\documentclass[a4paper,11pt]{article}
\usepackage[utf8]{inputenc}
\usepackage{graphicx}
\usepackage{geometry,mathrsfs}
\usepackage{amsmath}
\usepackage{homepackalllocal}

\usepackage{hyperref}
\usepackage{tikz}
\usetikzlibrary{shapes.misc}
\usetikzlibrary{plotmarks}
\usepackage[numbers]{natbib}
\usepackage{multicol}

\renewcommand{\b}{\mathbf}

\newcommand{\newtitle}[1]{
	\vspace{14pt} 

	\rm\large{\bfseries\boldmath #1 } \\
	\unboldmath
	}

\begin{document}
\rm\large \bfseries\boldmath%
\begin{center}
Modeling nonlinear wave-body interaction with the Harmonic Polynomial Cell method combined with the Immersed Boundary Method on a fixed grid.
\vspace{0.15cm}

\rm \underline{Fabien Robaux}$^1$, Michel Benoit$^2$\\
\rm\small{$^1$}
\rm\small{Aix-Marseille Univ. (AMU). Institut de Recherche sur les Phénomènes Hors Equilibre (IRPHE), UMR 7342}\\
\rm\small{(AMU, CNRS, Centrale Marseille), 13013 Marseille, France.}
\rm\small{\emph{fabien.robaux@irphe.univ-mrs.fr}}\\
\rm\small{$^2$}
\rm\small{Centrale Marseille \& IRPHE, 13013 Marseille, France.}
\rm\small{\emph{michel.benoit@irphe.univ-mrs.fr}}
\end{center}
\newtitle{Introduction}
To model the propagation of large water waves and associated loads applied to offshore structures, scientists and engineers have a need of fast and accurate models.
A wide range of models have been developped in order to predict wave-fields and hydrodynamic loads at small scale, from the linear potential boundary element method to complete CFD codes, based on the Navier-Stokes equations. 

\begin{figure}[htbp!]
	\begin{center}
	\begin{tikzpicture}
		[scale=0.90,
		ghost/.style={mark=square*, draw=black,fill=black!10, line width=0.5, mark options={fill=black!10}},
		 ghostEx/.style={shape=rectangle, draw=black, line width=0.5},
		 fluid/.style={shape=rectangle, draw=black,fill=blue, line width=0.5,mark=square*,mark options={fill=blue}},
		 marker/.style={cross out, minimum size=0.1, draw=black, line width=0.5},
		 interp/.style={shape=circle, draw=black, line width=1},
		 inactive/.style={shape=rectangle, draw=black,fill=black, line width=0.5,mark=square*,mark options={fill=black}, only marks},
		 neumann/.style={shape=rectangle, draw=black,fill=green, line width=0.5,mark=square*,mark options={fill=green}}]

		\draw (-0.4,1.2) grid (15.4,-4.4);
	%
		\draw [domain=0:9, samples=10] plot[fluid] (\x, 0);
		\draw [domain=9:15, samples=7] plot[fluid] (\x, -1);
		\draw[domain=0:15, samples=16, smooth] plot[mark=x, mark color=green] (\x, {0.7*sin(20*\x)+0.1});
		\draw [domain=0:10, samples=11] plot[ghost] (\x, 1);
		\draw [domain=10:15, samples=5] plot[ghost] (\x, 0);
		\node[draw] (P) at (10,1) {};
		\node[draw] (S) at (9,0) {};
		\draw[->, >=latex] (P) to (S); 

		\newcommand{\interpolation} 
		{(3,-1)(4,-1)(5,-1)(5,-2)(5,-3)(5,-4)(4,-4)(3,-4)(3,-3)(3, -2)};
		\draw plot[only marks,mark=*,mark options={fill=red}] coordinates {\interpolation};
		\draw[domain=-180:180, samples=11, color=green] plot[neumann] ({0.9*cos(\x)+4},{0.9*sin(\x)-2.3});
		\draw[domain=-180:180, samples=11, color=green] plot ({1.5*cos(\x)+4},{1.5*sin(\x)-2.3});
		\draw[domain=-180:180, samples=11, color=green] plot[mark=*] ({2.1*cos(\x)+4},{2.1*sin(\x)-2.3});
		\foreach \k in {0,36,...,324}
		{\draw[color=green]  ({0.9*cos(\k)+4},{0.9*sin(\k)-2.3}) -- ({2.1*cos(\k)+4},{2.1*sin(\k)-2.3});}

		\draw plot[inactive] coordinates {(4,-2)(4,-3)(11,1)(12,1)(13,1)(14,1)(15,1)};

		 \matrix [draw,above left, fill=white, row sep=-0.5 mm] at (current bounding box.south east) {
		  \node [ghost,label=right:\scriptsize{Ghost Points}] {}; \\
		  \node [fluid,label=right:\scriptsize Centers of cells used for ghost points] {}; \\
		  \node [marker,label=right:\scriptsize Markers on the free surface] {}; \\
		  \node [interp,label=right:\scriptsize Interpolation points from other mesh] {}; \\
		  \node [inactive,label=right:\scriptsize Inactive points] {}; \\
		  \node [neumann,label=right:\scriptsize Neumann BC points] {}; \\
		};
	\end{tikzpicture} 
	\caption{Schematic representation of the immersed free surface and immersed body}
	\label{schemaPoints}
	\end{center}
\end{figure}
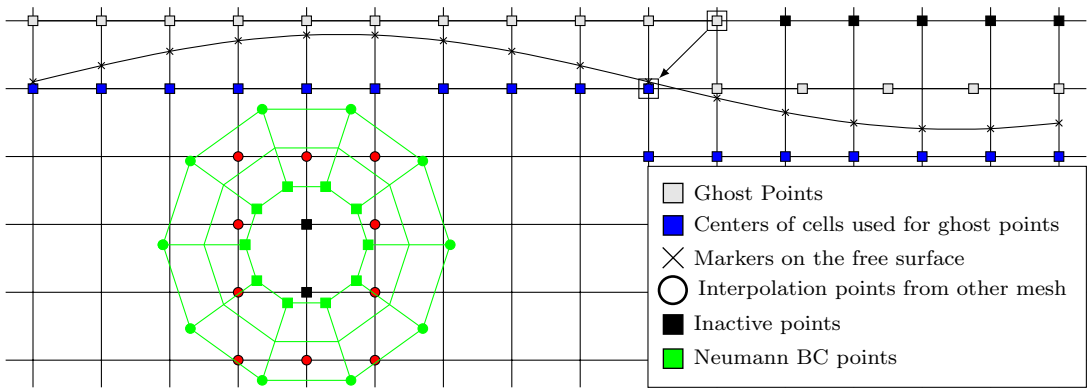

Although the latters are well adapted to solve the wave-structure interaction at small scale, their use is limited due to the computational cost of such models and numerical diffusion.
Alternative approaches, capturing the nonlinear effects, are thus needed.
\citet{HPC:shao2014harmonic} proposed an innovative technique, called "harmonic polynomial cell" (HPC) method to tackle this problem.
This approach is implemented and tested in 2 dimensions $(x,z)$, first on a standing wave problem and then to evaluate the nonlinear forces acting on a fixed submerged cylinder.
\newtitle{Overview of the harmonic polynomial cell method (HPC)}
This method is based on the potential hypothesis, which assumes the irrotationality of the flow.
Viscous effects are also neglected. 
The complete description of the velocity field can thus be reduced to the knowledge of the potential scalar field $\b{v}=\grads (\phi)$, where $\phi$ satisfies the Laplace equation:
\begin{eqnarray}
\nabla^2 \phi=0 \label{eqlaplace}, \hspace{2cm} 	 -h(\b{x}) \leq z \leq \eta(\b{x},t)
	\label{laplaceeq}
\end{eqnarray}
To solve the potential problem, we use the HPC method introduced by \citet{HPC:shao2014harmonic}.
The fluid volume is discretized into overlapping cells.
In each cell, composed of 9 points, the potential is approximated as a weighted sum of the first harmonic polynomials, each of them being solution of \equref{laplaceeq}.
The method can be extended to 3D cases considering cubic-like cells with 27 nodes.
\begin{multicols}{2}
	In a given cell the potential is thus locally approximated as a linear combination of the 8 first harmonic polynomials ($f_j$), which are explicitely known ($1$, $x$, $z$, $xz$, $x^2-z^2$...).
	\begin{equation}
		\phi(\b x=(x,z))=\sum_{j=1}^{8}b_j f_j(\b{x}) \label{interpEq}
	\end{equation}
	Thus, applying this equation to each node of the cell, except the center, gives $\phi_i=\phi(\b{x_i}) =  b_j f_j(\b{x_i}) $ for $i=1..8$.
	\begin{center}
	\begin{tikzpicture}[scale=0.30]
		\coordinate (1) at (0., 0.) ;		
		\coordinate (2) at (5, -0.5);		
		\coordinate (3) at (10, 1.1);		
		\coordinate (4) at (0., 5.2);		
		\coordinate (9) at (5, 4.1);		
		\coordinate (5) at (10, 5.5);		
		\coordinate (6) at (0, 10.1);		
		\coordinate (7) at (5., 10.4);		
		\coordinate (8) at (10., 9.1);
		\draw (1) node[left,below] {$1$};
		\draw (2) node[left,below] {$2$};
		\draw (3) node[left,below] {$3$};
		\draw (4) node[left,left] {$4$};
		\draw (9) node[left,below] {$9$};
		\draw (5) node[left,right] {$5$};
		\draw (6) node[left,above] {$6$};
		\draw (7) node[left,above] {$7$};
		\draw (8) node[left,above] {$8$};
		\draw (1) -- (2);		
		\draw (2) -- (3);		
		\draw (4) -- (9);		
		\draw (9) -- (5);		
		\draw (6) -- (7);		
		\draw (7) -- (8);		
		\draw (1) -- (4);		
		\draw (2) -- (9);		
		\draw (3) -- (5);		
		\draw (4) -- (6);		
		\draw (9) -- (7);		
		\draw (5) -- (8);				
		\coordinate (10) at (-2, 0.);
		\coordinate (11) at (0., -3);
		\coordinate (20) at (5, -3.);
		\coordinate (30) at (10, -3.);
		\coordinate (31) at (12, 1.1);
		\coordinate (40) at (-2, 5.1);
		\coordinate (50) at (12, 5.5);
		\coordinate (60) at (-2, 10.);
		\coordinate (61) at (0, 13.0);
		\coordinate (70) at (5.0, 13.0);
		\coordinate (80) at (10.0, 13.0);
		\coordinate (81) at (12.0, 10.0);
		\draw [dashed] (1) -- (10);
		\draw [dashed] (1) -- (11);
		\draw [dashed] (2) -- (20);
		\draw [dashed] (3) -- (30);
		\draw [dashed] (3) -- (31);
		\draw [dashed] (4) -- (40);
		\draw [dashed] (5) -- (50);
		\draw [dashed] (6) -- (60);
		\draw [dashed] (6) -- (61);
		\draw [dashed] (7) -- (70);
		\draw [dashed] (8) -- (80);
		\draw [dashed] (8) -- (81);
	\end{tikzpicture} 
	\end{center}
\end{multicols}
This local matrix (size 8x8 with $C_{ij}=f_j(\b x_i)$) can be inverted, such that the $b_j$ coefficients are obtained for the given local cell as $b_j =C^{-1}_{ji}\phi_i$.
Then injecting this result into the interpolation equation~\equref{interpEq}, a relation is obtained providing an approximation for the potential inside the cell using the potentials of the eight surrounding nodes: 
	\begin{equation}
		\phi(\b x)= \sum_{i=1}^{8} \left[ \left(\sum_{j=1}^{8} C^{-1}_{ji}  f_j(\b x) \right) \phi_i\right] \label{main}
	\end{equation}
Still, the potential at the center of the cell (node 9) has not been used to obtain~\equref{main}.
Thus, applying~\equref{main} at this node where $\b{x}=(0,0)$, a relation is found between the nine potentials of each cell. 
For each cell in the fluid domain, this equation is set in a global matrix. 
Thus this matrix contains at most 9 non-zero values in each row. 
For the boundary conditions, either a Dirichlet condition can be imposed or a Neumann condition is found by deriving the local expression~\equref{main}.
Following \citet{zakharov1968}, the kinematic and dynamic free surface nonlinear boundary conditions are formulated as:
\begin{eqnarray}
\eta_{t} & = & -\nabla \eta \cdot \nabla \tilde{\phi}+\tilde{w}(1+\nabla \eta \cdot \nabla \eta)  \label{eqzakharov1}\\
\tilde{\phi}_t & = & -g\eta-\frac{1}{2}\nabla\tilde{\phi}\cdot\nabla\tilde{\phi}+\frac{1}{2}\tilde{w}^2(1+\nabla\eta\cdot\nabla\eta), \label{eqzakharov2}
\end{eqnarray}
where  $\tilde{\phi}(x,t)=\phi(x,\eta(x,t),t)$, and $\tilde{w}(x,t)=\frac{\partial \phi}{\partial z}\big| _{z=\eta}$ is the vertical velocity at the free surface. The gradient of $\eta$ is computed by finite difference of arbitrary order. Others spatial derivatives can be computed locally by deriving the local expression for $\phi$~\equref{main}. Moreover the Runge-Kutta method at order 4 with constant time-step is used to integrate (Eq. \ref{eqzakharov1}-\ref{eqzakharov2}) in time.%
%
\newtitle{Immersed boundary method (IBM) on a fixed grid}
In a first step of this study, the HPC method was implemented using deforming boundary fitted grids, but a lack of stability was denoted.
The results were found to be highly dependent on the mesh deformation method used, due to the difficulty to invert the local geometry matrix at some time-steps and for particular lay-outs of the cell.
Recently, \citet{HPC:ma2017local} pointed out that the method HPC is much more efficient when using fixed perfectly-squared cell.
In order to work with regular fixed grids, an IBM strategy is implemented, following \citet{HPC:hanssen2017free}. 
In this method, the free surface is discretized with markers, evenly spaced and positionned at each vertical intersection with the background fixed grid (See schematic representation \normalfigref{schemaPoints}).
In order to close the system, each point above the free surface - denoted ghost-points - must have a dedicated equation in the global matrix.
Those equations are the local expressions \equref{main} applied at the marker position in a given cell (given $C_{ij}$).
A choice must be made when two ghost-points appear to be linked to the same maker in the same cell.
The second closest cell whose center is inside the fluid domain is thus chosen to prevent singularity, see arrow in \normalfigref{schemaPoints}.

To allow the background mesh to be composed of square cells, and still have the possibility to include fixed or moving objects, a boundary fitted overlapping grid is also added following \citet{HPC:ma2017local}.
In this method an immersed boundary-fitted mesh is constructed. 
Local expressions are also solved on this complementary mesh and added to the global matrix to solve both problems (on the background and immersed mesh) simultaneously. 
Communication is set and solved in the global matrix through interpolations using \equref{main} at the extreme points of both the background mesh and the immersed mesh, ensuring the continuity of the solution and the boundary condition on the body.
\newtitle{ Results on two selected test cases}
\rm\large{\emph{Standing wave freely evolving in a closed square domain}}\\
This first test case is compared to the results from the method of \citet{tsai1994numerical}. 
It consists in a standing wave of wavelength $\lambda=64m$, steepness $H/\lambda=0.1$ in a square domain with water depth $h=\lambda$.
The initial conditions for $\eta$ are set according to the results of \citep{tsai1994numerical} with a phase such that the velocity field is null.
The wave is freely evolving for 9 periods, and the error on $\eta$ is computed at each time step with an $L_2$ norm, and normalized with the wave height:
$	err=\frac{\sqrt{\sum_p (\eta_p-\eta_{th})^2}}{H}$.

\fig[Normalized $L_2$ error on $\eta$ - Standing wave]{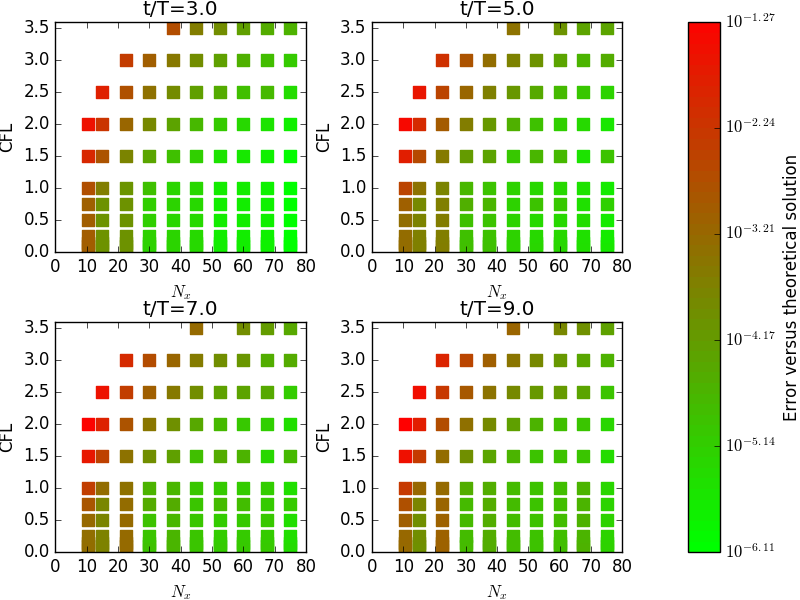}[0.6]

\dblfig[convergence in mesh refinement]{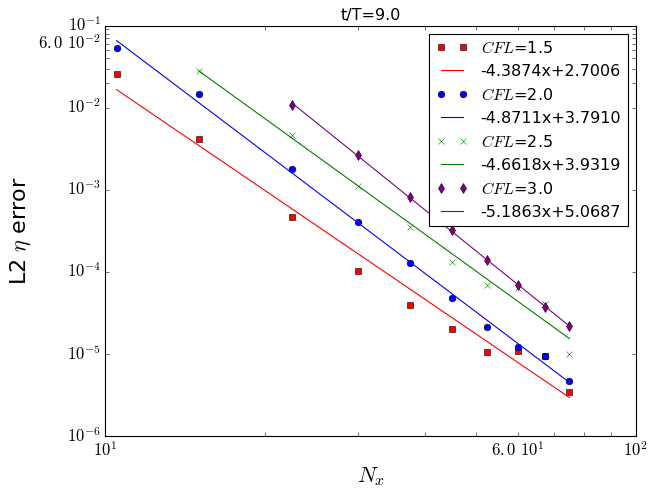}[convergence in time refinement]{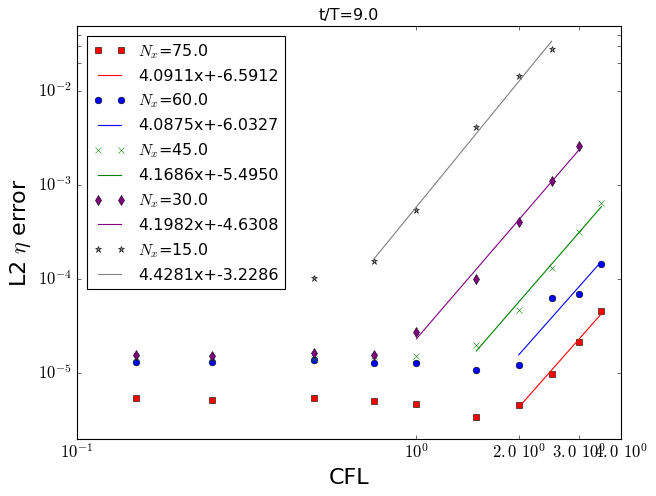}[Convergence in time and space]

This normalized $L_2$-error is plotted at four different times in \figref{ErrorTotStanding} as a function of the number of nodes per wavelength ($N_x$) and the Courant Friedrich Levy (CFL) number, defined as the radio of the number of time-steps per period divided by $N_x$.
Results are promising, with errors down to $10^{-6}$. 
Convergence properties with mesh and time-step refinement are also investigated and both are found to be with an order close to 4 as expected (\figref{Error2DconvxStanding}). 
\\
\\
\\
\\
\rm\large{\emph{Fixed horizontal immersed cylinder}}\\
\unboldmath
The second test case is based on the work of \citet{chaplin1984nonlinear},
which consists in a fixed horizontal cylinder, slightly immersed, in regular waves of period $T=1s$. 
The depth of the cylinder center is $d_c=0.101m$, for a radius $r=1/2 d_c$, and a total depth of $d=0.85m$.
This problem is numerically difficult to solve as the cylinder is close to the free surface, involving a small water gap to be meshed, of height $\approx \lambda/30$.
The incident waves are generated using the stream function theory.
The mean value and the amplitudes of the first 3 harmonics of the vertical load on the body are compared to numerical results from \citet{guerber2011modelisation}, computed with a non-linear boundary element method for different Keulegan-Carpenter numbers ($K_c$). 
	\fig[Different harmonics of the vertical load on the cylinder. Current results (dots) compared to numerical simulations from \citep{guerber2011modelisation} (lines with crosses).]{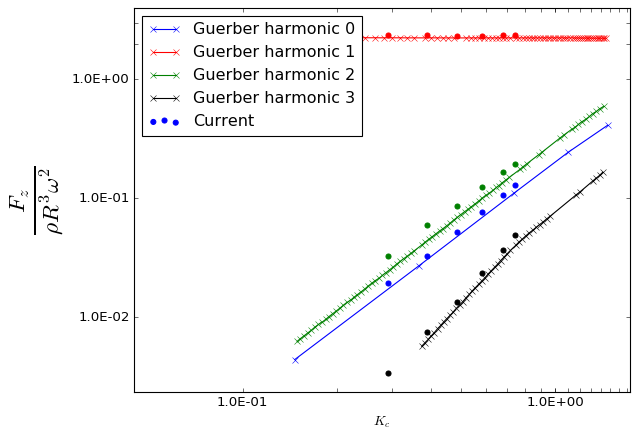}[0.6]

As presented in \figref{Error2DCyl}, results are in quite good agreement with \citet{guerber2011modelisation}, thought some differences are observed.
For smaller wave heights (ie, smaller $K_c$), the results found with a more classical HPC boundary fitted mesh approach were in good agreement with the litterature due to the small deformation of the mesh. 
Results from the two approaches will be compared and discussed during the workshop.
\bibliographystyle{abbrvnat}
\bibliography{ThesisBib}

\end{document}